\documentstyle[11pt,epsf]{article}
\def\beq{\begin{eqnarray}}
\def\eeq{\end{eqnarray}}
\def\eq{\end{eqnarray}}
\def\bea{\begin{eqnarray*}}
\def\eea{\end{eqnarray*}}

\def\centeron#1#2{{\setbox0=\hbox{#1}\setbox1=\hbox{#2}\ifdim
\wd1>\wd0\kern.5\wd1\kern-.5\wd0\fi 
\copy0\kern-.5\wd0\kern-.5\wd1\copy1\ifdim\wd0>\wd1 
\kern.5\wd0\kern-.5\wd1\fi}}
\def\ltap{\;\centeron{\raise.35ex\hbox{$<$}}{\lower.65ex\hbox{$\sim$}}\;}
\def\gtap{\;\centeron{\raise.35ex\hbox{$>$}}{\lower.65ex\hbox{$\sim$}}\;}

\def\singleandthirdspaced{\baselineskip=\normalbaselineskip\multiply
    \baselineskip by 130\divide\baselineskip by 100}

\newcommand{\newc}{\newcommand}
\newc{\qbar}{{\overline q}}
\newc{\Kahler}{K\"ahler }
\newc{\deltaGS}{\delta_{\rm GS}}
\begin{document}
\begin{titlepage}
\begin{flushright}
{\large hep-ph/0405159 \\ SCIPP-2004/15\\ SU-ITP 04/19\\
}
\end{flushright}

\vskip 1.2cm

\begin{center}

{\LARGE\bf Visible Effects of the Hidden Sector}

\vskip 1.4cm

{\large M. Dine$^a$, P.J.Fox$^a$, E. Gorbatov$^a$, Y. Shadmi$^b$, 
Y. Shirman$^c$ and S. Thomas$^d$}
\\
\vskip 0.4cm

{\it $^a$Santa Cruz Institute for Particle Physics,
     Santa Cruz CA 95064  } \\
{\it $^b$Physics Department,
     The Technion---Israel Institute of Technology, Haifa 32000, ISRAEL  }\\
{\it $^c$T-8, MS B285, LANL, Los Alamos, NM 87545 }\\
{\it $^d$Physics Department,
     Stanford University, Stanford, CA 94305}

\vskip 4pt \vskip 1.5cm

\begin{abstract}

The renormalization of operators responsible for soft 
supersymmetry breaking is usually calculated by starting at some 
high scale and including only visible sector interactions in the 
evolution equations, while ignoring hidden sector interactions.
Here we explain why this is correct only for the most trivial 
structures in the hidden sector, and discuss possible 
implications.  This investigation was prompted by the idea of 
conformal sequestering. 
In that framework hidden sector renormalizations by nearly 
conformal dynamics are critical.  In the original models of 
conformal sequestering it was necessary to impose hidden sector 
flavor symmetries to achieve the sequestered form.  We present
models which can evade this 
requirement and lead to no-scale or anomaly mediated boundary 
conditions; but the necessary structures do not seem generic.  
More generally, the ratios of scalar masses to gaugino masses, 
the $\mu$-term, the $B\mu$-term, $A$-terms, and the gravitino mass can be 
significantly affected. 

\end{abstract}

\end{center}

\vskip 1.0 cm

\end{titlepage}
\setcounter{footnote}{0} \setcounter{page}{2} 
\setcounter{section}{0} \setcounter{subsection}{0} 
\setcounter{subsubsection}{0}

\singleandthirdspaced

\section{Introduction--A Simple Argument for Conformal Sequestering}

Anomaly mediation \cite{anomalymediation,randallsundrum} is generally
regarded as an interesting, and possibly predictive, scheme for soft
supersymmetry breaking parameters.  In order that the anomaly-mediated
contributions dominate, however, it is necessary that the Kahler
potential has a particular, ``sequestered" form.  Some arguments for
how this structure might emerge for geometrical reasons in higher
dimensional constructions were given in \cite{randallsundrum}.
However, in many situations in string and M theory, where one might
have hoped to obtain geometric sequestration, the Kahler potential
does not have the required form \cite{dineetalam}.  The dominance of
anomaly mediated contributions in geometric realizations seems to
require that the underlying microscopic theory possess (non-abelian
discrete) symmetries acting on the fields of the visible
sector. However, such symmetries, if present, can already provide a
solution of the supersymmetric flavor problem.

Luty and Sundrum (LS) \cite{ls} have proposed an interesting
alternative way of realizing anomaly mediation in four dimensional
effective field theories called ``conformal sequestering," inspired,
in part, by warped constructions in higher dimensions.  They argue
that in theories in which the hidden sector is conformal over a range
of energies, the Kahler potential flows to the sequestered form. In
this note, we will review and slightly recast these arguments.  We
will stress that in the simplest realizations, as noted by LS,
additional continuous or discrete symmetries are required to achieve
the sequestered form.  Amusingly, the symmetries in this case must act
on hidden sector fields, as opposed to visible ones.  We will then
extend the analysis of LS to a class of models with gauge singlet
fields which require smaller continuous or discrete symmetries. We
also show that conformal sequestering may be achieved without any
hidden sector flavor symmetries, but only under rather exceptional
circumstances or with specially chosen couplings to hidden sector
singlets.

Apart from the question of anomaly mediation, the observations of Luty
and Sundrum lead us to focus on another issue: the role of hidden
sector interactions in the renormalization group evolution of
operators responsible for visible sector soft-breaking parameters.
The standard method to analyze theories with supersymmetry breaking in
a hidden sector has been the following. One supposes that the scale of
interactions which separates the visible from the hidden sector is
$M$.  (We have in mind theories where $M$ is large, such as theories
of gravity mediation, or gauge mediation with very massive
messengers).  One writes an effective action {\it for the visible
sector fields} at the scale $M$ treating the hidden sector fields as
non-dynamical background fields. This is a supersymmetric theory with
higher dimensional operators suppressed by the scale $M$ which couple
the visible and hidden sectors.  One then writes a set of
renormalization group equations for the higher dimensional operators
and evolves these into the infrared, {\it ignoring the hidden sector
dynamics}. Supersymmetry breaking $F$ and $D$ components of the
background hidden sector fields then give rise to visible sector soft
supersymmetry breaking parameters.

Stated in this way, it is clear that if hidden sector fields are 
treated as dynamical, then in addition to the usual contributions 
due to visible sector interactions, {\it hidden sector 
interactions can also, in general, give rise to appreciable 
renormalization group evolution of the operators responsible for 
soft supersymmetry breaking parameters.}  By considering the 
evolution of different types of soft masses, one can quickly 
derive the results of LS, and extend them to a broader class of 
models.  In this note, we will explain this phenomenon first for 
globally supersymmetric theories, then for locally supersymmetric 
theories.  We will illustrate the point first using a simple 
theory where it is easy to perform explicit computations, and we 
will then discuss gauge theories and conformal field theories, 
both weakly and strongly coupled.  We review the analysis of LS, 
stressing, as they noted,
that discrete symmetries are required in their models if anomaly
mediation is to dominate.  
We then go on to construct models with smaller and possibly no 
symmetry requirements.  
The general features of hidden sector renormalization on 
the operators responsible for direct gaugino masses, 
the $\mu$ and $B\mu$ terms, and $A$-terms  are also considered. 
The classes of models presented here with for conformal sequestering
with which do not require microscopic discrete symmetries 
are shown to yield either anomaly mediated or no-scale boundary 
conditions. 
We conclude with an assessment of anomaly 
mediation, arguing that, while it might occur in specially chosen 
models of geometric or conformal sequestering, it does not seem 
to be a generic outcome of any more fundamental framework with 
which we are currently familiar.
But more generally, hidden sector interactions 
can have an important impact on the spectrum of 
soft supersymmetry breaking parameters.

\section{Global Supersymmetry}

\subsection{Yukawa theories}
\label{yukawa} 
Consider a theory with two sets of fields, $Q$ and 
$T$, with superpotential 
\beq W = {1 \over 3!} \lambda T^3 \ ,
\eeq  
where throughout $T$ are understood to be hidden sector fields 
while $Q$ are visible sector fields. Suppose that at the high 
scale, $M$, there is also a higher dimension, Kahler potential 
coupling between the two sectors 
\beq {\cal L}_{TQ}=c \int d^4 
\theta T^\dagger T Q^\dagger Q\ . \label{ttqq} 
\eeq 
where $c(M) \sim 1 / M^2$. 

It is straightforward to compute the anomalous dimension of the 
operator in Eqn.~(\ref{ttqq}).  In terms of supergraphs, there is 
only one vertex diagram; one also has to include a wave function 
renormalization of the $T$ fields.  The result is
\begin{eqnarray}
\delta c &=& 3c {\lambda^2 \over 16 \pi^2} \ln(\mu/M)\ ,
\\
{d \over dt} c &=& 3 {\lambda^2 \over 16 \pi^2}c\ . 
\label{crenorm}
\end{eqnarray}
where throughout $t\equiv\log \mu$ is the logarithm of the 
renormalization scale. One can alternatively derive this result 
using arguments along the lines of \cite{nimaetal}.  Think of $Q$ 
as a background field, and define 
\beq 
Z_o = (1 + c Q^\dagger Q). 
\label{zo} 
\eeq 
At one loop, $Z$ is corrected; the $T$ propagator 
is proportional to $Z_o^{-1}$. So the self energy correction is 
proportional to $Z_o^{-2}$, and the $T$ effective Lagrangian is 
\cite{nimaetal}: 
\beq 
{\cal L}_T = Z_o\left(1-Z_o^{-3}{\lambda^2 
\over 16 \pi^2} \ln(\mu/M)\right)T^\dagger T 
\eeq 
which yields the same anomalous dimension for the operator ${\cal L}_{ttqq}$.

Let us briefly comment on the background field calculation above. 
Obviously, the wave-function coefficient $Z$ is not a measurable 
quantity, and physics can not depend on the change of 
normalization for kinetic terms. Thus we need to specify the 
effect of the redefinition (\ref{zo}) on the canonical coupling 
$\lambda_c=\lambda/Z^{3/2}$. To keep the physical coupling 
unchanged would require a non-holomorphic, field-dependent rescaling 
of the superpotential coupling $\lambda$. Since, in the limit of 
interest, $Q$ is a light dynamical field, such rescaling is not 
consistent with maintaining 
a manifestly supersymmetric Lagrangian.
Therefore, the superpotential coupling $\lambda$ (as 
opposed to the canonical coupling $\lambda_c$) must be held fixed.

Now we can consider the implications of this operator for 
supersymmetry breaking.  
Following LS we first imagine that the $F$-component of $Q$ has an
expectation value.  This gives rise to a soft mass for $T$, 
\beq 
m_t^2= c \vert F_Q \vert^2.  
\eeq
According to Eqn.~(\ref{crenorm}), 
\beq 
{d \over dt} m_t^2 = 3{\lambda^2 \over 16 \pi^2}m_t^2. 
\eeq 
This agrees with the running of the mass predicted in such a theory by
the standard analysis; see, e.g. \cite{mv}.

Now we turn to the case of interest and suppose that that $F_T \ne 0$
so that, at $M$, 
\beq 
m_q^2 = c \vert F_o\vert^2, 
\eeq 
where $F_o=F_T(M)$ is the $F$ term vev of $T$ at the messenger scale.
Notice that the soft mass is proportional to the vacuum energy in this
globally supersymmetric model.  As a result of the running of $c$, the
$Q$ mass evolves.  We need to also think about the renormalization of
$F_T$.  Generally $F_T$ will be a function of other fields in the
model.  If we assume that any non-perturbative renormalization of the
superpotential is unimportant, then we can write the Lagrangian as
before, but $F_T = Z_T^{-1/2} F_o$, where $F_T=F_T(\mu)$ is the $F$
term for the canonically-normalized field at the scale $\mu$.
Furthermore, $F_o$ should also be written in terms of the canonically
normalized fields at $\mu$.  So the $Q$ mass renormalizes with both
the running of $c$ and the running of the vacuum energy.  Relative to
the vacuum energy running of the soft mass is still given by a factor
$Z_T^{-3}$.  If the T sector has a natural scale (e.g. the scale of
supersymmetry breaking in the hidden sector, $M_{hid}$), then the
coefficient only runs down to this scale.

To illustrate this,
one can take an
O'Raiferataigh type model, with 
\beq 
W=  T(\lambda_o A^2-\xi_o^2) + m_o BA 
\eeq 
where $\lambda_o$, $m_o$ and $\xi_o$ are ``bare" 
parameters, i.e. parameters of the effective Lagrangian at the 
scale $M$.  
For large $m_o$, only $T$ develops an $F$-term, 
 $F_{To} = \xi_o^2$.
At lower scales, this expression can be written in terms of 
renormalized fields and parameters, where $\xi^2 = 
\xi_o^2/\sqrt{Z_T}$.  Now suppose that at the scale $M$, the 
theory contains the operator: 
\beq 
{\cal L}_{AQ}= a{1 \over M^2} 
A^\dagger A Q^\dagger Q.  
\eeq 
Since there is no $F$ term for $A$, this operator does
not give a soft mass for $Q$ at tree level.
However, at one loop, it induces the 
operator of Eqn.~(\ref{ttqq}), with coefficient,
\beq 
c = 2a{\lambda^2 \over 16\pi^2} \ln(\mu/M).  
\eeq 
(All of the quantities in this expression are renormalized quantities; note 
that the wave function of $T$ is not important at this order).
 
So this example is more dramatic: the soft mass vanishes at the 
scale $M$, but it is induced by the low energy dynamics of the 
hidden sector under renormalization group evolution to lower 
scales.  Again, note that this operator only runs down to the 
scale of the $T$ interactions (i.e. roughly to the scale $\xi$).
 
\subsection{Gauge theories}
 
The effects of hidden sector gauge interactions on the running of 
operators which couple the visible and hidden sectors can also be 
included. We wish to consider the renormalization of the operator 
of Eqn.~(\ref{ttqq}). From our analysis of the previous section, 
we can proceed very simply.  If we suppose first, that $F_Q \ne 
0$, the operator (\ref{ttqq}) generates a soft mass for the $T$ 
scalars.  In a pure gauge theory, the running of this soft mass 
is well known: 
\beq 
{d \over dt} (m^2)^i_j = {1 \over 16 \pi^2} (8 
\delta^i_j g^4 C(i){\rm Tr}(S(r)m^2)).
\label{twolooprunning} 
\eeq 
where $C(i)$ is the Casimir 
appropriate to the scalar in question; the trace is a sum over 
all of the matter fields, again with $S(r)$ being the appropriate 
character of the representation.

So we see that if we write the operator of Eqn.~(\ref{ttqq}) 
including the most general flavor structure 
\beq 
{\cal L}_{ttqq}= 
\sum_{ij;ab}c_{ij:ab}\int d^4 \theta T_i^{\dagger}T_j 
Q_a^{\dagger}Q_b \label{ttqqflavor} 
\eeq 
then, at weak coupling: 
\beq 
{d \over dt} c_{ij:ab} = {1 \over 16 \pi^2} (8 \delta_{ij} g^4 
C(i){\rm
  Tr}(S(r)c_{ij:ab}))
\label{cevolution} 
\eeq 
(the trace is on the $i,j$ indices).

This running can be derived more directly working with Feynman 
graphs.  One particularly simple approach is suggested by a 
discussion in LS. For simplicity of writing, consider a theory 
with vanishing superpotential, and $N_f$ vectorlike flavors.  
Study the flavor singlet operator: 
\beq 
{\cal L}_{ttqq}= c\sum_{ij}\int d4 \theta T_i^{\dagger}T_j Q_a^{\dagger}Q_a 
\eeq 
Various components of this operator may be considered. In 
particular, one of the terms in this expression is a product of 
currents, $j^\mu_T j_{\mu Q}$.  In the limit $c=0$, these 
currents are conserved.  Since the $Q$'s
are treated as non-interacting, it is tempting to say that
the operator appearing here is not 
renormalized.  But $j^\mu_T$ is in general an anomalous chiral 
current.  If the theory is regulated with a Pauli-Villars 
regulator of mass $M_{pv}$, then the renormalization can be 
determined by noting that because of the conformal anomaly the 
divergence of the current receives a contribution from the 
Pauli-Villars fields.  The matrix element of the corresponding 
density ($M_{pv} \bar \psi_{pv} \psi_{pv}$) is easily evaluated, 
and gives Eqn.~(\ref{cevolution}).

Note that if we consider an anomaly free hidden sector vector 
current, there is indeed no renormalization.  The same holds for 
operators which are flavor non-singlet in visible sector fields 
(as was discussed in \cite{ls}).

\section{Local Supersymmetry}

The analysis of theories with local supersymmetry is similar 
to the analysis of globally supersymmetric theories described above. 
We first consider the $\lambda T^3$ model in the locally 
supersymmetric case.  As an example, we suppose that the Kahler 
potential is that appropriate to ``minimal gravity mediation," 
i.e it is universal, with only quadratic terms in the chiral 
fields in the Einstein frame: 
\beq 
K = T^\dagger T + Q^\dagger Q\ . 
\label{quadratickahler} 
\eeq (where we have restricted our 
attention to a model with only these two fields).  We will ask, 
to order $\lambda^2$, what sorts of corrections are induced to 
$K$.  Again, we will see that the corrections account for the 
usual running of the soft masses.

The conventional Einstein frame form for the action is not the 
most convenient for this analysis.  Instead, the standard 
auxiliary field analysis \cite{cremmer,wb} may be done in 
the superconformal frame, where the Ricci scalar is multiplied by a 
function $\Phi$, of chiral fields.  $\Phi$ is related to the 
Kahler potential by: 
\beq 
K = -3 \ln( -\Phi/3)\ .
\eeq 
The Kahler potential of Eqn.~(\ref{quadratickahler}) 
corresponds to $\Phi$ which includes 
\beq \Phi \supset 
 T^\dagger T + Q^\dagger Q - {1 \over 6} (T^\dagger T +
Q^\dagger Q)^2\ .  
\eeq 
In this form, by focusing on particular terms 
in the component field action, it is easy to compute the 
renormalization of the various terms.  In particular, the 
term $T^\dagger T Q^\dagger Q$  is renormalized by the same 
diagrams (in terms of component fields) as in the globally 
supersymmetric case.  In other words, the function $\Phi$, at one 
loop in hidden sector fields, has the form: 
\begin{eqnarray}
\Phi & \supset & 
 (1 + \delta Z) T^\dagger T + Q^\dagger Q - 
  {1 \over 3} (T^\dagger T Q^\dagger Q) (1-2\delta Z) + \dots 
\\ & \supset & T^\dagger T + Q^\dagger Q - {1 \over 3} 
(T^\dagger T Q^\dagger Q)(1-3\delta Z) \ .
\end{eqnarray}
where 
\beq 
\delta Z = {\lambda^2 \over 32 \pi^2} \ln(\mu^2/M^2)\ , 
\eeq 
and in the second line we have rescaled (renormalized) the field $T$.  
The conventional Kahler potential is now given by 
\beq 
K 
\simeq T^\dagger T + Q^\dagger Q 
+(1-3 \delta Z)\, (T^\dagger TQ^\dagger Q)\ ..  
\eeq

One can also check that this works properly in the Einstein 
frame.  Some of the simplest terms to study are the quartic 
fermion couplings; it is easy to isolate them, compute their 
renormalization, and infer the correction to the Kahler 
potential.  If $F_Q \ne 0$, we again reproduce the standard 
running of the soft mass for $T$.  But again, if $F_T \ne 0$, we 
see that the soft masses for $Q$ run in the low energy theory.

\subsection{Are Hidden Sector Dynamics Important?}

In the next section, we will consider the proposal of Luty and 
Sundrum, that the supersymmetry-breaking sector is a strongly 
coupled field theory, conformal over a range of energies.  In 
this case, the effects of hidden sector renormalization of the 
quartic terms in the Kahler potential are profound.  First, 
however, we ask what one might expect the size of these effects 
to be in asymptotically free hidden sector theories which are 
weakly coupled at the messenger scale.  We will see that even in 
this case they can easily be of order one or larger.

As an example, we take our ``hidden sector" theory to be an 
$SU(N)$ gauge theory with $N_f$ flavors, $T_i$ and $\bar T_i$, 
and a singlet, $S$.  We take the singlet to couple to one of the 
flavors, with superpotential: 
\beq W= \lambda S T_1 \bar T_1. 
\eeq
The one-loop renormalization group equations for this theory are 
\beq
{d \lambda^2 \over dt} &=& -{c \lambda^2 \over 16\pi^2} g^2+ 2 {N+2 
\over 16 \pi^2} \lambda^4 \ ,\\
{dm_T^2 \over dt} &=& { \lambda^2 \over 16 \pi^2}
 \left( 2 m_S^2 + 4 N m_T^2 \right)\ ,\\
\frac{dg^2}{dt}&=&- {2 b_o \over 16\pi^2} g^4 \ ,
\eeq
where $m^2$ are the soft masses-squared, 
$c=4(N^2-1)/N$, and $b_o=3N-N_f$. 
The Yukawa  coupling has an attractive quasi-fixed point trajectory given by 
\beq 
{\lambda^2 \over g^2} = { c - 2 b_o \over 2(N+2)} 
\eeq 
Restricting to this trajectory for simplicity, the one-loop 
renormalization group equation for $m^2_T$ are easily solved 
\beq 
m_T^2(\mu)=m_T^2(M) \left( { g^2(\mu) \over g^2(M) } 
  \right)^{{N \over N+2}{2b_o - c \over b_o}}
\label{mTmu} 
\eeq

As before, we can interpret this as the renormalization of the 
coefficient of the operator $T^\dagger T Q^\dagger Q$, where $Q$ 
is a visible sector field, which for the purposes of the 
calculation above may be thought of as having an auxiliary field 
expectation value. To get a sense of how large the hidden sector 
renormalization effects can be, consider the exponent in 
(\ref{mTmu}). In the large $N$ limit at fixed $N_f$ it becomes 
${2 \over 3}$. So if the hidden sector gauge coupling becomes 
strong, the effects can be sizeable.  The expression (\ref{mTmu}) 
is only valid at one loop, but the loop expansion only begins to 
break down for $g^2(\mu) N \sim 16 \pi^2$, illustrating the 
possibility of very large effects.  Strong hidden sector dynamics 
is common in renormalizable theories of dynamically 
broken supersymmetry. So large renormalization effects should be 
typical of a wide classes of models.

Again, there are a variety of ways one might want to think about 
these results.  Instead of studying the running of operators due 
to hidden sector interactions starting at the scale $M$, one can 
alternatively (and equivalently) adopt the viewpoint that one 
should take boundary conditions for the renormalization group 
equations, not at scale $M$, but at the lower scale, $M_{hid}$.  
Of course, determining these conditions requires knowledge of the 
microscopic physics at $M$, and the full renormalization group 
evolution down to $M_{hid}$.

\subsection{Gravitino Mass} 

The gravitino mass is of course directly related 
to the scale of supersymmetry breaking in the hidden sector. 
Assuming the cosmological constant vanishes by a cancellation 
between auxiliary and superpotential expectation values, the 
gravitino mass squared is 
\beq
m_{3/2}^2 =  { 3|F|^2 \over M_p^2} 
\label{gravmass}
\eeq
where here $|F|^2$ is understood to be a sum over all 
renormalized auxiliary expectation values in the hidden sector. 
In the notation of section 2.1, the renormalized values 
of individual expectation values are 
$F = Z^{-1/2} F_o$. 
So effects of hidden sector renormalization 
appear also in the gravitino mass.  

In many models it is important to consider the
effects of hidden sector renormalization on visible sector 
soft parameters relative to the gravitino mass. 
This is particularly true for models of conformal sequestering
since anomaly mediated contributions to soft masses are 
suppressed by a loop factor compared with the gravitino mass.

\section{The Luty-Sundrum Model}

LS consider the case where the $T$ fields are part of a gauge 
theory which is nearly conformal over a range of scales.  The 
theories which they consider are required to have a global, 
non-abelian hidden sector flavor symmetry.  They present a 
sophisticated argument to determine the running of $c$, but the 
main features of the result are clear from our discussion above.  
Again suppose first, for illustrative purposes, that $F_Q \ne 0$. 
Then at $M$, the $T$ fields have a non-zero mass.  This will 
exhibit nearly power-law variation with scale since the hidden 
sector is nearly conformal. Thus the coefficient $c$ of the 
operator coupling the visible and hidden sectors does as well.  
In a strongly coupled theory, the anomalous dimension will be of 
order $1$.  Now, as in our perturbative discussion above, we can 
consider instead $F_T \neq 0$, to argue that the soft masses for 
the $Q$'s will exhibit power law renormalization since they arise from 
the same operator. 

Actually, the analysis of LS provides further important 
information: as in our perturbation theory discussion, it 
demonstrates that hidden sector flavor non-singlet operators have 
vanishing anomalous dimensions, while a certain hidden sector 
flavor-singlet operator has anomalous dimension equal to the 
first derivative of the beta function at the fixed point 
\cite{ls}. Since, by assumption, the fixed point is attractive, 
this ensures that the anomalous dimension of this operator is 
positive.  If the coefficients of the non-singlet operators 
vanish, then, in the supergravity context, our analysis of the 
previous section indicates what happens to $\Phi$: it behaves as: 
\beq 
\Phi \supset T^\dagger T + Q^\dagger Q + 
{\cal O}(e^{-b^\prime t})  T^\dagger T Q^\dagger Q \ .
\eeq 
This is the sequestered form of the Kahler potential. 
The soft masses arising from direct couplings between 
the visible and hidden sectors are exponentially suppressed
compared with the gravitino mass, and the 
anomaly-mediated contributions dominate.

In the case of a weakly coupled pure gauge theory (the Banks-Zaks 
theory \cite{bankszaks}, to be discussed in more detail below), 
we can use our earlier arguments to determine the behavior of 
the various operators. Thinking, again, of the $F_Q$ fields as 
the origin of soft masses for the $T$ fields, we can use the 
known results for this evolution.  At one loop, the anomalous 
dimensions of the soft masses (in the absence of gaugino masses 
and Yukawa couplings) vanish; at two loops they are given by 
Eqn.~(\ref{twolooprunning}), with $g$ the value of the coupling at 
the fixed point. Note that one sees immediately that the hidden 
sector flavor non-singlet operators (\ref{ttqqflavor}) have 
vanishing anomalous dimension.  The hidden sector flavor singlet 
operator (\ref{ttqqflavor}) with $i=j$ flows to zero in the 
infrared.

As we will see, even at weak coupling, it is easy to construct 
conformal field theories with much smaller global symmetries, in 
which almost all of the operator coefficients tend to zero.  This 
is demonstrated by simply introducing soft masses for the hidden 
sector fields and studying their evolution. Still, in the 
examples we study below, it is necessary either to impose some 
discrete symmetry or arrange rather special couplings in the 
hidden sector in order that anomaly mediation dominate.

In the next subsection, for completeness, we reproduce the LS 
argument for the anomalous dimension.  In the following 
subsection, we study weakly coupled fixed points, verifying the 
LS argument, but also developing theories in which all of the 
soft operators tend to zero.

\subsection{The Anomalous Dimension of the Soft Mass}

We have given a rather explicit derivation of the running of the 
soft masses in gauge theory.  However, there is another argument 
which generalizes straightforwardly to strongly coupled theories. 
The basic strategy, following Arkani-Hamed et al \cite{nimaetal}, 
is to think of $Z$ as a vector superfield, whose highest 
component is the soft mass.  This is most easily done for the 
flavor singlet case.  We will return to the non-singlet case 
shortly.  One wants to determine the $Z$-dependence of the 
effective action by studying the renormalization group evolution 
of the lowest component of $Z$, and then use this to infer the 
behavior of the highest component.  LS consider a situation where 
$T T^\dagger QQ^\dagger$ should be thought of as giving a 
correction to $Z_T$.  They then use this sort of argument to 
infer soft masses for $Q$.  But this can just as well be used to 
infer soft masses for $T$ (due to $F_Q F_Q^\dagger$).

In a gauge theory, naively, there is only a uniform rescaling of 
the kinetic terms, i.e. they are always proportional to $Z_o$, 
the $Z$ factor at the cutoff scale.  This is because factors of 
$Z$ in vertices are always compensated by factors of $Z$ in 
propagators. Thinking of $Z$ as including the corrections to the 
Kahler potential, as in Eqn. (\ref{zo}), this would seem to say 
that the soft masses are not renormalized, since the kinetic 
terms and soft mass terms would be rescaled by the same factor.  
On the other hand, we know that at two loops, there is a 
renormalization of the soft masses.

The resolution of this question was provided by Arkani-Hamed et 
al. As Shifman et al explained originally~\cite{nkvz}, the 
$Z$-dependence of the action is more complicated, because 
introduction of a regulator introduces new $Z$ dependence.  This 
can be described by relating the physical and holomorphic gauge 
coupling constants through the relation (valid at all scales) 
\beq 
\label{ghol} {8 \pi^2 \over g^2(\mu)} = {8 \pi^2 \over g_h^2(\mu)} 
-N_c \ln g^2(\mu) - N_f \ln Z(\mu) 
\eeq 
This relation reconciles 
the exact NSVZ beta function with the requirement that the coupling 
constant in the Wilsonian action evolves only at one loop due to 
the holomorphy. 
Once wave-function and holomorphic gauge couplings are promoted 
to background superfields, the physical coupling must also be 
treated as a real superfield. Similarly to the case of the Yukawa 
theories discussed in section (\ref{yukawa}), the holomorphic 
gauge coupling must be held fixed while the perturbation in the 
Kahler potential is turned on. This requirement together with 
Eqn.~(\ref{ghol}) determines the relation between the perturbation 
of $\tau={8\pi^2}/{g^2}$, and the perturbation of the 
wave-function, $\delta Z$: 
\beq 
\label{taupert} \delta \tau = \frac{\delta \tau_{hol} - N_f 
\ln (1+\delta_Z/Z)}{1 - N/\tau}\ .
\eeq 
It is easy to see that $\delta \tau$ does not evolve
  at one loop.

Let's use this to recover the result for the evolution of the soft 
masses quoted by Martin and Vaughn \cite{mv}.  The usual solution 
of the renormalization group equation for $Z(\mu)$ is: 
\beq 
Z(\mu) = Z(M) e^{\int_{g(M)}^{g(\mu)}{dg^\prime \gamma(g^{\prime}) 
\over \beta(g^\prime)}} \ .
\eeq 
Writing 
\beq 
\beta(g) = -{g^3 b_o \over 16 \pi^2}\ , 
~~~~~\gamma(g) = c{g^2 \over 16 \pi^2} \ ,
\eeq 
this gives: 
\beq 
Z(\mu) = Z(M) \left ({g_p^{-2}(M) \over 
g_p^{-2}(\mu)} \right )^{-c/2 b_o} \ .
\eeq
This is the standard evolution of $Z$ in terms of the gauge 
coupling of the {\em perturbed} theory, $\tau_p=\tau+\delta \tau$. 
Similarly to the Yukawa theory, we can expand the above equation 
in terms of the perturbation $\delta \tau$. The dependence on the 
perturbation arises at order $g^4$ -- as expected it is a two loop 
effect. Expanding the above in $\delta \tau$ and using the Eqn. 
(\ref{taupert}) gives the correct beta function for the soft mass. 
This argument is clearly another version of the calculation we 
described in the previous section.

\subsection{Weakly Coupled Conformal Field Theories}

Let's study the problem of soft masses at a Banks-Zaks fixed 
point.  We will start with a conventional analysis, using the 
formulas in, say, \cite{mv}. The first two beta function 
coefficients are: 
\beq
b_o &=& -(3N-N_f){g^3 \over 16\pi^2}\ ,\\ 
b_1 &=& 
-\Big[6N^2-2N N_f -4N_f{N^2 -1 \over 2N}\Big]\,
\left({g^2 \over 16 \pi^2}\right)^2 g. 
\eeq
Requiring that the beta function vanish
gives, to this order, 
\beq 
N_f=3N-\epsilon\ , ~~~~~
{g^{*2} \over 16\pi^2}={\epsilon \over 6N^2}\ . 
\eeq 
Here we will think of $N$ as extremely large, and of $\epsilon$ 
as an integer of order $1$. There will be corrections to this 
result at higher orders in $g$. In particular, at higher orders, 
$g^*$ {\it is scheme dependent}. 

Note that near the fixed point, 
\beq 
g = g^* + \delta ~~~~~\beta= 
\beta^\prime(g)\delta = b\delta = {\epsilon^2 \over 3N^2} \delta 
\eeq So \beq \delta(t) = \delta(0)e^{-bt}. \label{deltarunning} 
\eeq 
Similarly, 
\beq 
\gamma \approx c = -{g^2 N \over 16 \pi^2} = c 
+ d~\delta, 
\eeq 
where $c = -{N \over 6} \epsilon$; $d=-{2N \over 
g}{g^2 \over 16\pi^2}.$ The renormalization group equation for $Z$ 
then leads to: 
\beq 
Z(\mu) =Z(M)\, e^{ct+{d \over b}(\delta(t)-\delta(0))}. 
\eeq

Indeed, for the Banks-Zaks case the LS argument gives the correct 
anomalous dimension for the soft masses, $\gamma = \beta^\prime$ 
at the fixed point.  The coefficient out front also works out 
correctly, i.e. one gets the solution with the correct boundary 
condition to the renormalization group Eqn.

Following LS, we can obtain a general expression for the 
anomalous dimension, valid at strong coupling.  LS ask how the 
effective action depends on $Z_o$. Near the textbook solution of 
the renormalization group equation to first order in $\delta 
\tau$, the distance of $\tau$ from the fixed 
point, is: 
\beq 
Z(t) = (Z(0)+\delta_Z(0)) e^{\gamma^* t + {\gamma^\prime \over
    \beta^\prime} (\delta \tau(t) -\delta \tau(0))} 
\eeq 
where $\gamma^\prime$ and $\beta^\prime$ are the derivatives 
of the $\gamma$ and $\beta$ functions with respect to $\tau$.  
~From the NSVZ formula for the $\beta$ function, we also have: 
\beq 
\beta^\prime = N_f{\gamma^\prime \over (1 - N/\tau)}.  
\eeq 
Exploiting the Eqn. (\ref{taupert}) we obtain 
\beq 
\delta \tau(t) = {\delta \tau_{hol} + N_f \ln(1+\delta_Z(0)/Z(0)) \over (1-N/\tau)} 
e^{\beta^\prime t}.  
\eeq 
Substituting back into the expression 
for $Z(t)$, the factor of $\delta_Z(0)/Z(0)$ cancels, and we are left with: 
\beq 
Z(t) = Z(0)e^{\gamma^*t}e^{{\ln(1+\delta_Z(0)/Z(0))~ \delta 
\tau(t)}} 
\eeq 
which reproduces the result for the running of the soft mass of the
Banks-Zaks fixed point.

\subsection{Purging Flavor Symmetries}

As discussed by Luty and Sundrum, only a particular operator 
(\ref{ttqqflavor}) with $i=j$ which is a flavor singlet under 
hidden sector flavor symmetry is renormalized in the theory 
discussed above. The model has an $U(N_f) \times U(N_f)$ hidden 
sector flavor symmetry at the classical level.  All but one of 
the operators $T_j T_j^\dagger$ contain, as their $\theta \bar 
\theta$ components, non-anomalous conserved currents, and are 
thus not renormalized.  The one exception is the anomalous axial 
$U(1)$.  As discussed above, it is easy to derive the 
renormalization of this operator at two loops from the anomaly.  
If one is to obtain anomaly mediation, it is then necessary to 
suppress all the operators (\ref{ttqqflavor}) which are flavor 
non-singlet in the hidden sector  at the high scale.  In 
\cite{ls}, this is achieved through intricate hidden sector 
discrete symmetries. 

But our earlier discussion suggests that if we can construct a 
hidden sector CFT without flavor symmetries, it should be 
relatively easy to suppress all operators (\ref{ttqqflavor}) 
connecting the hidden and visible sectors. 
As discussed below, it is rather difficult
to construct conformal models with only gauge-non-singlet fields and
without continuous non-$R$ global symmetries. But this can be
accomplished if hidden sector gauge singlet fields are included with
appropriate superpotential couplings.  As an example, consider
supersymmetric $SU(N)$ QCD with $N_f$ flavors and $N$ colors in the
conformal window ${3 \over 2} N \leq N_f < 3N$. We add $N_f$ singlets,
$S_i$, with tree level superpotential 
\beq 
W= \lambda \sum_{i=1}^{N_f} S_i T_i \bar T_i \label{Wmeson} 
\eeq 
which also results in an
interacting conformal theory. To achieve such a coupling, one can
impose an $S_{N_f}$ permutation symmetry, or -- even better -- simply
note that the equations for the fixed point have such a symmetry at
one loop.  Before writing down explicit formulas note that $\lambda$
has an attractive fixed point, this can be easily verified at weak
coupling.  As a result, ignoring contributions from hidden sector 
gaugino masses, many soft masses run to zero (this is a corollary 
of a theorem of Nelson and Strassler \cite{nelsonstrassler}).  
However, the model has $N_f$ vector-like conserved $U(1)$'s 
corresponding to the Cartan sub-algebra of the vector $SU(N_f)$ 
and the overall vector $U(1)$. Correspondingly, these $N_f$ 
linear combinations of masses (or more precisely, coefficients of 
quartic operators) do not run.

This models respects an $S_{N_f}$ permutation symmetry at the 
fixed point, as well as a $Z_2$ charge conjugation symmetry which 
exchanges $T_i \leftrightarrow \bar T_i$. If we suppose that this 
discrete $S_{N_f} \times Z_2$ symmetry is exact in the microscopic 
theory, then the dangerous operators would be absent, and anomaly 
mediation would dominate.  The requisite number of symmetries may 
be reduced further by introducing additional singlets with 
different couplings to the quarks.  For example, adding $N_f-1$ 
additional singlets, $S^{\prime}_i$, with tree level 
superpotential 
\beq 
W=\lambda^{\prime} \sum_{i=1}^{N_f-1} S_i^{\prime} T_i \bar{T}_{i+1} 
\label{Wmesonperm} 
\eeq 
leaves the overall vector $U(1)$ as the only non-anomalous non-$R$
continuous symmetry. However, for $SU(N)$ gauge group with hidden
sector flavors, the most general singlet superpotential couplings to
mesons which preserve conformal invariance can not eliminate the
overall vector $U(1)$ symmetry. So in this class of models at least an
exact discrete $Z_2$ symmetry in the hidden sector is required in
order to suppress all the dangerous operators (\ref{ttqqflavor}) and
obtain conformal sequestering.

The requirement of an exact $Z_2$ symmetry can be relaxed in 
special circumstances. For $SU(N)$ gauge group with flavors this 
possibility arises only for the special cases of $N=4$, $N_f=8$ 
and $N=6$, $N_f=9$. In these cases baryon and anti-baryon 
operators with any flavor structure are exactly marginal and may 
be added to the superpotential without spoiling conformal 
invariance. For example, for the $SU(4)$ with $N_f=8$ theory the 
superpotential 
\beq 
W = \lambda^{\prime \prime} \left( T_1 T_2 T_3 T_4 + 
    \bar{T}_1 \bar{T}_2 \bar{T}_3 \bar{T}_4 \right)
\eeq 
breaks the overall vector $U(1)$. With inclusion of a 
sufficient number of singlets with couplings to mesons such as 
(\ref{Wmeson}) and (\ref{Wmesonperm}), all continuous non-$R$ 
symmetries are then broken. For $SO(N)$ and $SP(N)$ gauge groups 
with flavors, all the flavor symmetries are axial. So for these 
two classes of theories with sufficiently many singlets, 
superpotential couplings of the form (\ref{Wmeson}) and 
(\ref{Wmesonperm}) are alone sufficient to eliminate all 
continuous non-$R$ symmetries without requiring any microscopic 
discrete symmetries. 

In the absence of gauge singlet fields it is rather difficult to 
construct a single gauge group CFT which does not possess non-$R$ 
continuous symmetries at the infrared conformal fixed point.  A 
number of requirements must be met to obtain such a theory. 
First, in the theory with vanishing superpotential there must 
exist at least one chiral operator which transforms under each 
non-anomalous continuous non-$R$ symmetry. Second, at least one 
operator for each symmetry must be exactly marginal with $U(1)_R$ 
charge $R=2$. This is a rather non-trivial requirement since it 
is often the case that operators which break global symmetries 
are in fact (marginally) irrelevant at a fixed point so that 
there are enhanced global symmetries in the infrared. And third, 
each such operator which breaks each symmetry must appear in the 
superpotential of the interacting theory {\it after} including 
the most general set of field redefinitions.  Taken together 
these requirements are rather non-trivial and we expect that they 
are all satisfied only in rather exceptional circumstances. 
However, one CFT which does appear to satisfy all these 
requirements is supersymmetric $SU(4)$ QCD with $N_f=8$ flavors. 
As mentioned above the baryon and anti-baryon chiral operators of 
this theory are exactly marginal for any flavor structure, so 
adding these to the superpotential can break flavor symmetries
without spoiling conformal invariance.  
For example, the superpotential couplings 
\beq 
W = \lambda \sum_{i=1}^8 ~\left( T_i T_{i+1} T_{i+2} T_{i+3} +  \bar{T}_i 
\bar{T}_{i+1} \bar{T}_{i+2} \bar{T}_{i+3} \right) 
\eeq 
break the 
$SU(8) \times SU(8) \times U(1)$ non-anomalous symmetry to 
$U(1)^{12}$. And the superpotential couplings 
$$
W = \lambda^{\prime} \sum_{i=1}^4 ~\left( T_i T_{i+2} T_{i+3} 
T_{i+4} + T_i T_{i+1} T_{i+3} T_{i+4} + T_i T_{i+1} T_{i+2} 
T_{i+4} \right.
$$
\beq ~~~~~~~~~~~~~~~~~~~~~~~~~~ \left. + \bar{T}_i \bar{T}_{i+2} 
\bar{T}_{i+3} \bar{T}_{i+4} + \bar{T}_i \bar{T}_{i+1} 
\bar{T}_{i+3} \bar{T}_{i+4} + \bar{T}_i \bar{T}_{i+1} 
\bar{T}_{i+2} \bar{T}_{i+4} \right) \eeq 
break these remaining symmetries. So in this theory with at least
these superpotential couplings, all the operators (\ref{ttqqflavor})
are suppressed by hidden sector conformal dynamics and conformal
sequestering results without any microscopic discrete symmetries.

We leave to future work the construction of all CFT's without 
continuous non-$R$ symmetries; but conclude that without rather 
exceptional circumstances and/or  without specially chosen 
couplings possibly to gauge singlet fields, such CFT's do not 
appear to be generic.

\section{Gaugino Masses and Other Soft Parameters} 

Hidden sector renormalization can in principle affect any of the
operators responsible for visible sector soft supersymmetry breaking
parameters, including not only scalar masses but gaugino masses and
the $\mu$- and $B\mu$-terms.  Gaugino Majorana masses require that a
gauge singlet field acquire an auxiliary expectation value.  This can
be the conformal compensator of the supergravity multiplet (as is
assumed to dominate in anomaly mediation) or gauge singlet fields
directly in the hidden sector.  Below we consider the effects of
hidden sector renormalization on gaugino mass contributions from the
latter type of singlets.  This is particularly relevant to the class
of nearly conformal models introduced in the previous subsection in
which hidden sector gauge singlets are introduced in order to relax
the requirement of hidden sector flavor symmetries.

The effective operators which yield gaugino 
masses from direct couplings to the hidden sector
are of the form 
\beq
c_{\lambda} \int d^2 \theta ~S 
  W^{\alpha} W_{\alpha}
\label{gauginoop}
\eeq
where $S$ is a hidden sector singlet, 
$W_{\alpha}$ is a visible sector superfield strength, 
and $c_{\lambda}(M) \sim 1/M$. 
An auxiliary expectation value $F_S \neq 0$ 
gives a gaugino mass at renormalization scale $\mu$ of
\beq
m_{\lambda}(\mu) = c_{\lambda}(\mu)F_S(\mu) 
\label{gauginomass}
\eeq
Since the operator (\ref{gauginoop}) is linear 
in the singlet, $S$, the gaugino mass (\ref{gauginomass}) 
is renormalized by hidden 
sector dynamics {\it only} through the wave function 
of $F_S$. From the discussion of section 2.1, 
the renormalized auxiliary 
expectation value is 
$F_S(\mu) = Z_S^{-1/2} F_{S,o}$ where $F_{S,o} = F_S(M)$ 
and $Z_S(M) \equiv 1$. 
So the gaugino mass is 
\beq
m_{\lambda}(\mu) = c_{\lambda}(\mu) 
  {F_{S,o} \over Z_S^{1/2} } 
\label{gauginor}
\eeq
Written in this way it is clear that the coefficient 
$c_{\lambda}$ is renormalized only by visible sector 
dynamics. 
As an aside, note that unitarity requires the renormalized 
wave function factor of any gauge singlet field to be 
greater than one, $Z_S \geq 1$. 
The renormalized auxiliary 
expectation value appearing in the gaugino mass (\ref{gauginor})
also appears in the gravitino mass (\ref{gravmass}) which 
is likewise 
renormalized by hidden sector dynamics only through wave function 
effects. 
So 
we see that 
if supersymmetry breaking is dominated 
by gauge singlet auxiliary expectation values then 
{\it hidden sector interactions do not modify the  
operators responsible for direct gaugino soft masses
relative to the gravitino mass}. 
This is in contrast to the operators (\ref{ttqq}) responsible 
for scalar soft masses which can be either be suppressed or 
enhanced by hidden sector renormalization effects, 
as illustrated in sections 2.1 and 3.1 respectively. 
This is also in contrast to visible sector renormalization
of the operator (\ref{gauginoop}) 
which is either positive or negative depending 
on whether the gauge group is asymptotically 
free or not. 

It is also possible to consider the effects of hidden sector
renormalization on the operators responsible for the 
Higgs sector $\mu$- and $B\mu$-terms. 
The superpotential $\mu$-term, $W = \mu H_u H_d$, can arise
from hidden sector supersymmetry breaking with an operator 
of the form 
\beq
c^{\prime } \int d^4 \theta ~ S^\dagger H_u H_d
\label{muop}
\eeq
where $c^{\prime}(M) \sim 1/M$. 
Like the gaugino mass operator (\ref{gauginoop}) this 
operator is only linear in the hidden sector singlet $S$. 
So from the discussion above it follows that 
if supersymmetry breaking is dominated 
by gauge singlet auxiliary expectation values then 
{\it hidden sector interactions do not modify the operator 
responsible for the $\mu$-term relative to the gravitino mass}.
The soft supersymmetry breaking Lagrangian $B \mu$ term, 
${\cal L} = B \mu H_u H_d$, can arise from an operator
of the form 
\beq
c^{\prime \prime} \int d^4 \theta ~ T^{\dagger} T H_u H_d
\label{Bmuop}
\eeq
where $c^{\prime \prime}(M) \sim 1/M^2$.
This operator is bi-linear in fields which either
transform under hidden sector gauge interactions or are 
gauge singlet, and is of the general 
form of the operators (\ref{ttqq}) responsible for soft 
scalar masses. 
So hidden sector interactions can in principle 
either suppress or enhance the operators responsible 
for the direct $B \mu$-term relative to the gravitino mass. 

Finally, it is also possible 
to consider the effects of hidden sector
renormalization on the operators responsible for
scalar tri-linear $A$-terms. 
These terms arise from operators of the form 
\beq
c_A \int d^4 \theta~ S ~\Phi \Phi \Phi 
\label{Aop}
\eq 
where here $\Phi$ are understood to be 
visible sector chiral multiplets and 
$c_A(M) \sim 1/M$. 
The specific prefactors of the operators (\ref{Aop})
may depend on the underlying theory of flavor. 
Like the gaugino mass (\ref{gauginoop}) and $\mu$-term 
(\ref{muop}) these operators are also linear in the hidden sector
singlet $S$. 
So from the discussions above if follows that 
if supersymmetry breaking is dominated 
by gauge singlet auxiliary expectation values then 
{\it hidden sector interactions do not modify the operators 
responsible for $A$-terms relative to the gravitino mass}.

There are clearly various possibilities for how hidden sector 
interactions might renormalize the different operators which couple
to the visible sector. 
So small, moderate, or even large ratios of different
types of soft supersymmetry 
breaking parameters could result, depending on details
of the hidden sector. 

One possibility for large ratios of direct soft parameters
to the gravitino mass is the 
Luty-Sundrum class of models \cite{ls}.
The hidden sector is assumed to contain only 
gauge non-singlet fields and flow to an attractive conformal 
fixed point.  
As discussed in section 4.3, with appropriately chosen discrete symmetries 
or a conformal sector without non-$R$-continuous symmetries, 
the direct scalar masses are vanishingly small. 
From the discussion above we see that the direct $B\mu$- term 
is also vanishingly small.  
And since there are no hidden sector singlets, the direct 
gaugino masses, $\mu$-term, and $A$-terms
also vanish at leading order. 
Anomaly-mediated contributions to both scalar and gaugino 
masses then dominate. 
The $\mu$- and $B\mu$-terms must be generated by additional interactions
within the visible sector.
This class of models, however, suffers the problem  
of a tachyonic slepton endemic to all models of anomaly-mediated 
supersymmetry breaking. 
This unacceptable feature might be cured by yet additional visible 
sector interactions. 

An interesting ratio of direct soft parameters arises for 
the models of the section 4.3 
with a hidden sector with an attractive conformal fixed point, 
and with sufficiently many singlets with special couplings 
which break all the continuous non-$R$-symmetries. 
In this case, if hidden sector singlets dominate the 
supersymmetry breaking auxiliary expectation values, 
then from the discussion above, for general operators 
coupling the visible and hidden sectors the gaugino masses,
$\mu$-term, and $A$-terms are of order the gravitino mass, 
$m_{\lambda} \sim \mu \sim A \sim m_{3/2}$, 
while the direct scalar masses and $B\mu$-term are vanishing 
small, $m_{\phi}^2, ~B\mu \ll m_{3/2}^2$. 
The leading direct contributions to the scalar masses 
are anomaly-mediated and therefore loop-suppressed 
compared with the gauginos masses.
So this class of models yields an interesting set 
of no-scale boundary conditions.  
Scalar masses are generated mainly from renormalization 
group effects in the visible sector proportional to 
gaugino masses and possibly $A$-terms. 
Aside from the $H_u$ mass, all scalar masses 
are then non-tachyonic.   
And the $B\mu$-term is likewise generated by 
visible sector renormalization effects proportional 
to the product of gaugino masses and $\mu$-term. 
Note that these models yield a phenomenologically viable 
set of boundary conditions for the correct pattern of 
electroweak symmetry breaking 
without postulating additional 
visible sector interactions to modify soft parameters. 
However, since the $A$-terms transform under visible sector
flavor symmetries, this class of models does not necessarily 
solve the supersymmetric flavor problem without an underlying 
theory of flavor.

\section{Conclusions}

From the discussions of the previous sections, we draw two lessons.  
The first lesson 
has to do with the idea of conformal sequestering.  
In general, in a conformal field theory, we 
might expect large corrections to the operators which generate 
soft masses.  LS have provided an elegant analysis which relates 
the anomalous dimensions of these operators to the derivative of 
the beta function at the fixed point, in a certain class of 
theories.  But the present analysis also indicates that generic 
conformal theories possess continuous non-$R$ global symmetries 
and therefore do not lead, in any general way, to sequestering.  
Many of the dangerous operators which could violate visible 
sector flavor symmetries for example, simply do not run.  One 
must suppose that these operators are absent from the microscopic 
theory.  But in a theory like string theory which contains 
gravity, there are no continuous global symmetries, so one must 
suppose that the absence of these terms results from hidden or 
visible sector discrete symmetries, with a very special 
structure.  Indeed, in the original Luty-Sundrum model, an 
intricate set of hidden sector discrete symmetries are imposed in 
order to obtain sequestering. However, we have seen that it is 
fact possible to obtain conformal theories without non-$R$ 
continuous symmetries, but only in rather exceptional cases or 
with specially chosen couplings to gauge singlet fields. In 
the end, it is not clear to us that this can be viewed, in any 
sense, as a generic outcome to be expected of some sort of more 
fundamental theory.  Only a full survey of such theories might 
give an indication.

We view this result as a positive one.  In general, in thinking 
about Beyond the Standard Model physics, there are few limits to 
our speculations.  But the realization that geometrical 
separation does not lead, 
generically,
to sequestering 
suggests that anomaly mediation is not a likely outcome of any 
fundamental theory, such as 
known string models.
Conformal sequestering, as originally proposed, requires the same sorts
of discrete symmetries as geometric sequestering,
if it is to dominate.  We have seen that it may be possible
to achieve such sequestering without discrete symmetries,
but this phenomenon does not seem generic.

The second lesson is quite general: the evolution of 
the operators responsible for soft parameters 
cannot be determined 
without knowledge of hidden sector dynamics.  It is not enough to 
start with knowledge of the operators which couple the hidden and 
visible sectors at, say, the Planck scale and evolve these down 
to low energies using only visible sector dynamics. Hidden sector 
dynamics will also lead to evolution of the operators. If the 
hidden sector is, as one might imagine, strongly coupled in order 
to break supersymmetry, these effects could be quite large.  The 
Luty-Sundrum model is an extreme example of this phenomenon.  
Alternatively, and equivalently, it is necessary to provide 
boundary conditions for the renormalization group equations for 
the operators, not at the scale $M$, but at the scale at which 
the hidden sector decouples, $M_{hid}$.

It should be remarked that the importance of hidden sector 
dynamics discussed in this paper is an issue for the Wilsonian 
approach to renormalization, where by definition dynamics in both 
the hidden and visible sectors are integrated out scale by scale. 
In this approach the hidden sector can affect physics over all 
scales down to $M_{hid}$ and must be taken into account in the 
running of the non-renormalizable operators which couple the 
hidden and visible sectors. However, since the hidden and visible 
sectors are by definition not coupled through renormalizable 
interactions, the effects of each sector on the running of 
non-renormalizable operators factorize. It is then possible, in 
principle, to integrate out the hidden sector completely and 
arrive at a 1PI effective action for visible sector fields alone.
In this case, the presence of non-trivial hidden sector dynamics 
can be subsumed into the boundary conditions for the visible 
sector soft parameters {\it at the messenger scale} $M$. In this 
approach, hidden sector dynamics does not directly affect the 
renormalization group running of visible sector parameters below 
the messenger scale $M$. But, as discussed here and as is 
apparent in the Luty-Sundrum model, it can have an important 
impact on the relative magnitude of different soft parameters at 
the high scale.
In particular ratios of scalar masses to gaugino masses, 
the $\mu$ term, the $B \mu$ term, $A$-terms, and the gravitino mass, 
can be significantly affected.

\noindent {\bf Acknowledgements:}

\noindent We thank T. Banks, T. Bhattacharya, A. Friedland, M.  Luty,
E. Poppitz, R. Sundrum and J. Terning for conversations.  The work of
M.D., P.F. and E.G. was supported in part by the U.S.  Department of
Energy. M. Dine also acknowledges support from the U.S.-Israel
Binational Science Foundation.  Y.~Shadmi thanks UCSC-SCIPP and the
SLAC theory group for hospitality while this work was initiated.  The
research of Y.~Shadmi is supported in part by the Israel Science
Foundation (ISF) under grant 29/03. The research of Y. Shadmi and
Y. Shirman was supported in part by the United States-Israel
Binational Science Foundation (BSF) under grant 2002020.  Y.~Shirman
thanks UCSC-SCIPP and Physics Department at Technion for hospitality
during the work on this project.  Y.~Shirman is a Richard P. Feynman
fellow at Los Alamos National Laboratory and is supported by the US
Department of Energy.  The work of S.T.  was supported in part by the
US National Science Foundation under grant PHY02-44728.


\end{document}